\documentclass[12pt]{article}
\begin{document}
\title{{\bf Topology of the Universe  }}

\author{W{\l}odzimierz Piechocki \vspace{0.5cm}\\
{\small Theory Division, So{\l}tan Institute for Nuclear Studies}\\  
{\small Ho\.{z}a 69, 00-681 Warsaw, Poland; e-mail: piech@fuw.edu.pl}} 

\date{}
\maketitle

\begin{abstract}
\noindent
General relativity is unable to determine the topology of the Universe. 
We propose to apply quantum approach. Quantization of 
dynamics of a test particle is sensitive to the spacetime topology. 
Presented results for a particle in de Sitter spacetimes  favor 
a finite universe. 


\end{abstract}

Specification of the topology of the Universe is necessary for 
the correct interpretation of 
cosmological observational data. The problem is that in case our 
space is multiply connected or/and compact one can see multiple 
images of cosmic objects by looking along different null geodesics. 
The topology is also crucial for the description of the birth of the 
Universe. The resolution of the problem of Universe's large-scale
homogeneity may depend on the specification of the topology as well.
Finally, determination of the topology would answer the old question of 
a finite or infinite Universe[1,2].

The Einstein equations being partial differential equations cannot 
specify the topology of spacetime but only its local properties. 
In fact, there are always  many topologically distinct universe models 
consistent with a given local geometry. The determination of the 
topology requires some understanding beyond general relativity.

We propose to apply quantum approach. It turns out that imposition of 
a quantum formalism upon the classical dynamics of a test particle 
is sensitive to the choice of the spacetime topology [3,4,5]. In what 
follows we shall demonstrate this dependence for very simple cases 
of a test particle in two de Sitter spacetimes having the same 
local geometries but completely different topologies.  The two 
considered spacetimes are $V_p := (R^1 \times R^1,g)$ and 
$V_h :=(R^1 \times S^1,g)$. 
In both cases the metric $g$ is defined by the line-element
\begin{equation}
ds^2=dt^2- e^{2mt}\;dx^2,
\end{equation}
where $m^2=\Lambda /3$ and $\Lambda >0$ is a cosmological constant.

\noindent 
In case of $V_p$ the space has the topology  $R^1$, so it is non-compact 
and simply connected. The $V_h$ is defined to be a one-sheet  
hyperboloid embedded in 3d Minkowski space. Since the space  
of $V_h$ has the topology  $S^1$, it is compact and multiply connected. 
There exists an isometric immersion of $V_p$ into $V_h~$ [6]. 
It is defined by the mapping
\begin{equation}
V_p \ni (t,x)\longrightarrow (y^0,y^1,y^2) \in V_h,
\end{equation}
where
\[ 2my^0:=e^{mt}-e^{-mt}+(mx)^2~e^{mt}, \]
\[ 2my^1:=e^{mt}+e^{-mt}-(mx)^2~e^{mt}, \]
\[ y^2:= x e^{mt} \]
and one has 
\[ (y^2)^2+(y^1)^2-(y^0)^2=(1/m)^2.\]

Action integral describing a relativistic test particle of mass 
$m_0$ in gravitational field is defined by 

\begin{equation}
S:=-m_0\int d\tau \,\sqrt{g_{\mu\nu}(x^0(\tau),
x^1(\tau))\;\dot{x}^\mu(\tau)\dot{x}^\nu(\tau)},
\end{equation}
where $\tau$ is an evolution parameter, $(x^0,x^1)=(t,x)$ are 
spacetime coordinates, $\dot{x}^\mu := dx^\mu/d\tau ~~(\mu =0,1), 
$ and $g_{\mu \nu}~ (\mu,\nu =0,1)$ are metric tensor components.

\noindent It is known that Hamilton's principle applied to (3) leads 
to geodesic equations. The set of all solutions to the geodesic 
equations defines an extended phase-space of the system. The physical 
phase-space is its subset [4]. It consists of all time-like geodesics 
which are admitted by the spacetime topology and which satisfy a gauge 
condition resulting from reparametrization invariance of the action 
integral (3). 

Let us find a general form of a geodesic curve of our test particle. 
Instead of solving the geodesic equations directly we use local symmetry 
of the system and apply Noether's theorem. Obtained expressions for dynamical 
integrals depend explicitly on particle spacetime coordinates and can 
be converted into expressions for spacetime coordinates in terms of 
dynamical integrals. This way one finds a general form of 
geodesic curve of a test particle. In both cases, $V_p$ and $V_h$, 
infinitesimal symmetry transformations lead to three dynamical 
integrals satisfying $sl(2,R)$ algebra [4,5]. Since the three integrals  
are constants of motion, it is natural to choose them  to represent 
classical observables of the system. The integrals are functionally 
dependent since the action (3) is reparametrization invariant. This 
gauge condition constrains the observables to a one-sheet hyperboloid 
(OSH). Each point of OSH defines a particle trajectory. However, the 
sets of all possible trajectories ( the physical phase-space ) of the 
two considered systems are quite different. The topology of $V_h$ admits 
trajectories defined by any point of OSH. Thus the symmetry group 
of the system is $SO_\uparrow (2,1)~$ [5]. In contrary to the $V_h$ 
case, topology of $V_p$ restricts particle trajectories to the 
plane. As a result, one infinite curve of points on OSH is not available 
for particle dynamics and the physical phase-space is now isomorphic 
to $R^2$. Consequently, some global transformations generated by 
$sl(2,R)$ algebra are not well defined and $SO_\uparrow (2,1)$ 
is no longer the symmetry group of the system ( only translations and 
dilatations are the symmetry transformations ) [4].

Difference between $V_p$ and $V_h$ systems shown at the 
classical level lead to dramatic difference at the quantum level. 
Quantization in considered cases means finding representation of 
$sl(2,R)$ algebra of observables on a Hilbert space having at least 
two properties: (i) it is a self-adjoint representation 
and (ii) it is a unitary representation of the symmetry group of the 
corresponding classical system.

In the case of $V_h$ system there exists the representation of $sl(2,R)$ 
algebra on the Hilbert space $L^2(S^1)$ which can be lifted to the 
unitary representation of $SO_\uparrow (2,1)$ group [5]. The $V_p$ 
case is quite different. The representation of $sl(2,R)$ algebra is 
parametrized by continuous parameter $\alpha \in S^1.$ Each choice 
of $\alpha$ defines a unitary representation of the universal 
covering group $\widetilde{SL}(2,R)$ on the Hilbert space $L^2([0,2\pi])$ 
and representations corresponding to different $\alpha 's$ are 
unitarily nonequivalent [4].

In summary, the two considered systems are locally the same but they 
differ globally due to different topologies of corresponding spacetimes. 
In case spacetime has topology $R^1 \times S^1$ there is one-to-one 
correspondence between classical and quantum levels. The canonical 
quantization applied to the system having spacetime with topology 
$R^1 \times R^1 $ leads to infinitely many unitarily nonequivalent 
quantum systems. Therefore, presented results for the two `toy models' 
favor a  finite universe. 

It seems that our results can be generalized to the cases of a particle 
dynamics in 4d de Sitter spacetimes with $R^1 \times R^3$ and 
$R^1 \times S^3$ topologies, since the formulae for the line element (1) 
and the isometric immersion map (2) extend to 4d only by the increase of 
the number of space coordinates [6]. 
Presented method can be further extended to include sophisticated 
universe models with spacetimes defined locally by the Robertson-Walker 
metrics.

In conclusion, since quantum description of a system is 
 more fundamental than the classical one, compatibility 
of spacetime with quantization procedure might be used in the search 
for the topology of the Universe. 

We suggest to relate the universe models having topologies 
compatible with quantum dynamics of a test particle ( like $V_h$ case ) 
with the search for the topology  based on fluctuations 
in the cosmic microwave background [7].

\vspace{0.5 cm}
{\bf Acknowledgements}

The author thanks G. Jorjadze and A. Trautman for stimulating 
discussions.


\begin{thebibliography}{99}
\bibitem{1} J.-P. Luminet, G. D. Starkman and J. R. Weeks,  
            Scien. American  4 ( 1999 ) 68;  
            M. Lachieze-Rey and J.-P. Luminet, Phys. Rep.  254  
            ( 1995 ) 135.
\bibitem{2} Class. Quantum Grav.  15 ( 1998 ), No 9 ( special 
            issue featuring invited papers from the Topology of the 
            Universe Conference, Cleveland, Ohio, Oct. 1997 ). 
\bibitem{3} G. Jorjadze  and W. Piechocki, Class. Quantum Grav. 
             15 ( 1998 ) L41.
\bibitem{4} G. Jorjadze and W. Piechocki, Theor. Math. Phys. 
             118 ( 1999 ) 183.
\bibitem{5} G. Jorjadze  and W. Piechocki, Phys. Lett. B ( 1999 ), 
            accepted for publication (  gr-qc/9811094 ).
\bibitem{6} E. Schr\"{o}dinger, Expanding Universes,  
             Cambridge University Press, Cambridge, 1956 ;
            S. W. Hawking and C. F. R. Ellis, The Large Scale 
            Structure of Space-Time,  Cambridge University Press, Cambridge, 
            1973.
\bibitem{7} J. Levin,  J. D. Barrow,  E. F. Bunn and  J. Silk, 
            Phys. Rev. Lett.  79 ( 1997 ) 974;
            J. Levin, E. Scannapieco, G. de Gasperis, J. Silk and  
            J. D. Barrow,  Phys. Rev.  D58 ( 1998) 123006; 
            N. J. Cornish,  D. Spergel  and  G. Starkman, Phys. Rev. Lett. 
            77 ( 1998) 215;  Phys. Rev.  D57 ( 1998 ) 5982;           
J.R. Bond,  D. Pogosyan  and  T. Souradeep, Class. Quantum Grav. 
            15 ( 1998 ) 2671.

\end{thebibliography}
\end{document}